\documentstyle[preprint,aps,eqsecnum]{revtex}
\begin{document}

\thispagestyle{empty}
\setcounter{page}{0}

\title{HARD DENSE LOOPS IN A COLD NON-ABELIAN PLASMA}

\author{Cristina Manuel}

\address{Center for Theoretical Physics,
Laboratory for Nuclear Science, and Department of Physics\\
Massachusetts Institute of Technology, Cambridge,
Massachusetts 02139}

\maketitle

\thispagestyle{empty}
\setcounter{page}{0}

\begin{abstract}
$\!\!$Classical transport theory is used to study the response
of a non-Abelian plasma at zero temperature and high chemical
potential to weak color electromagnetic fields. In this article
the parallelism between the transport phenomena occurring in a
non-Abelian plasma at high temperature and high density is
stressed. Particularly, it is shown that  at high densities it is also
possible to relate the transport equations to the zero-curvature
condition of a Chern-Simons theory in three dimensions,
even when quarks are not considered ultrarelativistic.
The induced color current in the cold plasma
can be expressed as an average over angles, which
represent the directions of the velocity vectors of quarks
having Fermi energy. From this color current it is possible
to compute $n$-point gluonic amplitudes, with arbitrary $n$.
It is argued that these amplitudes are the same as the ones
computed in the high chemical potential limit of QCD, that
are then called hard dense loops.
The agreement between the two different formalisms is checked
by computing the  polarization tensor of QED due to finite
density effects in the high density limit.
\end{abstract}

\vfill

\hfill\break
\hbox to \hsize{CTP\# 2497\hfil December 1995}
\vskip-12pt
\eject

\baselineskip=15pt
\pagestyle{plain}
\section{INTRODUCTION}
\label{sec1}

Quantum Chromodynamics (QCD) undergoes a phase transition
 at high temperature and/or high density \cite{Kapusta}.
 Above a critical temperature and critical
chemical potential, quarks and gluons are no longer confined.
Those extreme  regimes are expected to be found in nature in
 some cosmological and  astrophysical
settings ({\it e.g.}, in the interior of neutron stars) or in heavy
ion collisions.
These are the regimes of QCD that we are going to discuss here.

As is well known now,  naive perturbative analysis of high
temperature QCD fails completely. This was realized when physical
quantities,
such as the gluonic damping rate, were found to be
gauge dependent when computed following the standard rules of quantum
field theories at finite temperature. The connection between expanding
in loops and expanding in the gauge coupling constant $g$ is not valid in
this regime. As was realized by Braaten and Pisarski \cite{BP1}, as well as  by
Frenkel and Taylor \cite{FT}, there are one-loop corrections, the hard thermal
loops (HTLs), which are as important as tree amplitudes, and they have to be
resummed and included consistently in all computations to non-trivial
order in $g$. Those hard thermal loops arise only for {\it soft} external
momenta ($\sim g T$) and {\it hard} internal loop momentum ($\sim T$),
where $g \ll 1$, and $T$ denotes the plasma temperature.
There is an infinite set of HTLs which possess very interesting
properties, such as obeying QED-like Ward identities and being
gauge independent \cite{FT}, \cite{BP2}.

The hard thermal loop resummation techniques that were proposed in
\cite{BP1} were successful in providing
one-loop gauge independent physical answers.
Much related work has been done since their discovery (see ref.
\cite{Nair}, \cite{Kobes}
for a review and references), but here we shall only review different
approaches to HTLs.

Taylor and Wong \cite{TW} were able to construct an
effective action for hard thermal loops just by solving the gauge invariance
condition imposed on that effective action. Efraty and Nair \cite{EN1},
\cite{EN2} have
related that condition to
the zero curvature equation of a Chern-Simons theory in (2+1) dimensional
space at zero temperature, providing therefore
a non-thermal framework to study a thermal effect.
This identification
has been used by  Jackiw and Nair \cite{JN}
to derive a non-Abelian generalization of the Kubo formula.

 Other derivations of the effective action of hard thermal
loops can be found in the literature. Blaizot and Iancu \cite{BI} could
extract the hard thermal loop effective action by studying the
truncated Schwinger-Dyson hierarchy of equations, after performing
a consistent expansion in the gauge coupling constant and obtaining
quantum kinetic equations for the QCD induced color current.
Jackiw, Liu and Lucchesi \cite{JLL} have also shown that HTLs can be derived
from the Cornwall-Jackiw-Tomboulis composite effective action \cite{CJT},
after requiring its stationarity.

A different derivation of the effective action of HTLs has
been given in \cite{KLLM}, which doesn't make use of quantum
field theory, as opposed to all previous approaches.
Hard thermal effects are exclusively due to thermal fluctuations,
and that is why a classical formalism to describe them was developed
in \cite{KLLM}. Just by writing the classical transport equations
for non-Abelian particles \cite{EH}, and using an approximation
scheme that respects the non-Abelian gauge symmetry of the transport
equations, the effective action of the infinite set of HTLs of QCD
could be found.

A similar situation may be expected to arise for QCD at high density
and zero temperature \cite{Braaten}. Actually, HTLs were also studied when a
chemical potential was included for quarks, and it was concluded that
the only effect of the chemical potential was modifying the Debye mass
 by a term proportional to the chemical potential \cite{BP3}, \cite{ViTo}.
 Therefore,  at very high density or chemical potential $\mu$
 and zero temperature,
one also may expect that  naive one-loop computations are incomplete, as
one-loop diagrams with {\it soft} ($\sim g \mu$) external momenta,
and  quarks running inside the loop with Fermi energy, are comparable to
tree amplitudes, and therefore they would have to be resummed.
We shall call those diagrams hard dense loops (HDLs) in analogy
to the thermal case.

The purpose of this article is to give a derivation of the hard dense
loops of gluonic amplitudes of QCD by using the classical
transport formalism. We expect that quantum field computations
of QCD at high chemical potential
will reproduce the transport results, exactly as happens in the
high temperature case, although we do not attempt here
to check complete agreement between these two different formalisms.

Although we study only gluonic amplitudes in this article,
it should be expected that
hard dense loops also arise for Feynman diagrams with external quark
legs. We do no take into consideration those diagrams in the present
article, and refer to a recent publication \cite{BO} which considers
the one-loop self-energy of the electron
for QED at finite density for those kinds of HDLs.

We have structured this article as follows. In Section II we
review briefly the classical transport theory for a non-Abelian
plasma. In Subsection III A we study how a cold quark plasma initially
in equilibrium reacts to weak external color electromagnetic
fields, and write the
transport equation that the induced color current density obeys.
 The parallelism
between the high temperature and high density cases is clearly established,
and we also comment about the connection between the transport equation
and the Chern-Simons eikonal.
In Subsection III~B we solve the transport equation in momentum space,
using recent results of ref. \cite{Zhang}.
In Subsection III C we give the polarization tensor in the quark plasma, and
prescribe how to derive higher-order point amplitudes.
As a check of the agreement between the transport approach and the
high chemical potential limit of QCD, we compute the
polarization tensor of QED due to finite density effects in Section IV,
and from it we extract the HDL corresponding to that graph,
which coincides with the one of QCD, up to some color factors.
Complete agreement is found between the two different computations.
On this result, we base our expectation
that complete agreement for higher $n$-point gluonic amplitudes
should also be true.
Finally, we present our conclusions.

\section{CLASSICAL TRANSPORT THEORY FOR A NON-ABELIAN
PLASMA}
\label{sec2}

The classical transport theory for the QCD plasma was developed
in~\cite{EH} and further studied in~\cite{KLLM},
 and here we will briefly review it. Consider a particle
bearing a non-Abelian $SU(N)$ color charge $Q^{a}, \ a=1,...,N^2-1$,
traversing a worldline ${\hat x}^{\alpha}(\tau)$.  The Wong equations
\cite{Wong} describe the dynamical evolution of the
variables
 $x^{\mu}$, $p^{\mu}$ and
$Q^{a}$ (we neglect here the effect of  spin):
\begin{mathletters}
\label{wongeq}
\begin{eqnarray}
m\, {{d {\hat x}^{\mu}(\tau) }\over{d \tau}} & = & {\hat p}^{\mu} (\tau)
\ , \label{wongeqa} \\[2mm]
m\, {{d {\hat p}^{\mu} (\tau)}\over{d \tau}} & = &
 g\, {\hat Q}^{a} (\tau)\, F^{\mu\nu}_{a} ({\hat x}) \, {\hat p}_{\nu} (\tau)
\ ,\label{wongeqb}\\[2mm]
m\, {{d {\hat Q}^{a} (\tau) }\over{d \tau}} & = & - g\,
f^{abc} \, {\hat p}^{\mu} (\tau) \,
A^{b}_{\mu} ({\hat x}) \, {\hat Q}^{c} (\tau)\ , \label{wongeqc}
\end{eqnarray}
\end{mathletters}
$\!\!$where $f_{abc}$ are the structure constants of the group, $F_a ^{\mu\nu}$
denotes the field strength, $g$ is the coupling constant, and we
set $c= \hbar = k_B =1 $ henceforth.

The main difference between the equations of electromagnetism and the
Wong equations, apart from their intrinsic non-Abelian structure, comes
from the fact that color charges precess in color space, and therefore
they are dynamical variables. Equation (\ref{wongeqc}) guarantees
 that the color current associated to each colored particle,
\begin{equation}
j_{\mu} ^a (x) = g \int d \tau \,  {\hat p}_{\mu}(\tau)\,  {\hat Q}^a (\tau)
\, \delta^{(4)} ( x -
 {\hat x}(\tau)) \ ,
\label{punc}
\end{equation}
where $({\hat x}^{\mu}, {\hat p}^{\mu}, {\hat Q}^a)$ are solutions of the
equations of
motion, is covariantly conserved
\begin{equation}
 \left(D_{\mu} j^{\mu} \right)_a (x) =  \partial_{\mu} j^{\mu}_a  (x)
+ g f_{abc} A_{\mu}^b (x) j^{\mu c}  (x) = 0 \ ,
\label{covcons}
\end{equation}
 therefore preserving the consistency of the theory.

The usual $(x,p)$ phase-space is thus enlarged to $(x,p,Q)$ by
including  color degrees of freedom for colored particles.
Physical constraints are enforced by inserting delta-functions in
the phase-space volume element $dx\,dP\,dQ$. The momentum
measure
\begin{equation}
dP = {{d^{4}p}\over{(2\pi)^{3}}}\,\,2\,\theta(p_{0})\,\,
\delta(p^{2} - m^{2})
\label{measurep}
\end{equation}
guarantees positivity of the energy and on-shell evolution. The color
charge measure enforces the conservation of the group invariants,
{\it e.g.}, for $SU(3)$,
\begin{equation}
dQ = d^8 Q\,\, \delta(Q_{a}Q^{a} - q_{2})\,\,
\delta(d_{abc}Q^{a}Q^{b}Q^{c} - q_{3}) \ ,
\label{measureq}
\end{equation}
where the constants $q_{2}$ and $q_{3}$ fix the values of the
Casimirs and $d_{abc}$ are the totally symmetric group constants.
The color charges which now span the phase-space are dependent
variables. These can be formally related to a set of independent
phase-space Darboux variables \cite{KLLM}. For the sake of simplicity, we will
use the standard color charges.

The one-particle distribution function $f(x,p,Q)$ denotes the
probability for finding the particle in the state $(x,p,Q)$.
In the collisionless case,  it evolves in
time via a transport equation
$ {{d f}\over{d \tau}} = 0$. Using the equations of motion~(\ref{wongeq}),
it becomes the Boltzmann equation \cite{EH}
\begin{equation}
p^{\mu}\left[{{\partial}\over{\partial x^{\mu}}}
- g\, Q_{a}F^{a}_{\mu\nu}{{\partial}\over{\partial p_{\nu}}}
- g\, f_{abc}A^{b}_{\mu} Q^{c}{{\partial}\over{\partial Q_{a}}}
\right] f(x,p,Q) = 0 \ .
\label{boltzmann}
\end{equation}

A complete, self-consistent set of non-Abelian Vlasov equations for
the distribution function and the mean color field is obtained by
augmenting the Boltzmann equation with the Yang-Mills equations:
\begin{equation}
[D_\nu F^{\nu\mu}]^a(x) = J^{\mu\, a}(x) =  \sum_{\rm species}\
 \sum_{\rm helicities}\
j^{\mu\, a}(x)\ ,
\label{yangmills}
\end{equation}
where the  color current $j^{\mu\,a}(x)$
for each particle species is
computed from the corresponding distribution function as
\begin{equation}
j^{\mu\,a} (x) = g\, \int dPdQ\ p^\mu Q^a f(x,p,Q) \ .
\label{cr5}
\end{equation}
Notice that if the particle's trajectory in phase-space would be known exactly,
then Eq. (\ref{cr5}) could be expressed as in Eq. (\ref{punc}).
Furthermore, the color current (\ref{cr5}) is covariantly conserved,
as can be shown by using the
Boltzmann equation \cite{KLLM}.

 The Wong equations (\ref{wongeq})
are  invariant  under the finite gauge
transformations (in matrix notation)
\begin{equation}
\label{gaugetrsf}
\bar{x}^{\mu}=x^{\mu}\ , \qquad
\bar{p}^{\mu}= p^{\mu}\ , \qquad
\bar{Q} =  U \,Q \,U^{-1}\ , \qquad
{\bar A}_\mu = U\,A_\mu \,U^{-1}-{1\over g}\,U\,
{\partial\over \partial x_\mu}\,U^{-1}\ ,
\end{equation}
$\!\!$where $U=U(x)$ is a group element.

It can be shown \cite{KLLM} that the Boltzmann equation (\ref{boltzmann})
 is invariant under the above gauge transformation if
the distribution function behaves as a scalar
\begin{equation}
{\bar f}({\bar x},{\bar p},{\bar Q}) = f (x,p,Q) \ .
\end{equation}
To check this statement it is important to note that under a gauge
transformation the derivatives appearing in the Boltzmann equation
(\ref{boltzmann}) transform as:
\begin{equation}
{\partial\over\partial x^\mu}=
{\partial\over\partial\bar{x}^\mu}
- 2 ~{\rm Tr}~ \Biggl([\ ({\partial\over\partial {\bar x}^\mu}U)
U^{-1}\ ,\  \bar{Q}\ ]
{\partial\over\partial\bar{Q}}\Biggr) \ , \qquad
{\partial\over\partial p^\mu}=
{\partial\over\partial\bar{p}^\mu} \ , \qquad
{\partial\over\partial Q}=
U^{-1} {\partial\over\partial\bar{Q}}U
\label{eq:gauge2c}\ ,
\end{equation}
that is, they are not gauge invariant by themselves. Only
the specific combination of the spacial and color derivatives
that appears in (\ref{boltzmann}) is
gauge invariant.

 The color current
(\ref{cr5}) transforms under (\ref{gaugetrsf}) as a gauge covariant vector:
\begin{equation}
{\bar j}^{\mu}({\bar x})= g\,\int dP\,dQ\,p^\mu\,U\,Q\,
U^{-1}\,f(x,p,Q)=U\,j^{\mu}(x)\,U^{-1}\ .
\end{equation}
This is due to the gauge invariance of the phase-space measure and to the
transformation properties of $f$.

\section{INDUCED COLOR CURRENT IN A COLD QUARK PLASMA}
\label{sec3}

\subsection{Transport equation for the color current}

In this section we study soft disturbances of
a completely degenerate quark plasma. We consider a quark plasma at
zero temperature and finite density which is initially in equilibrium.
In the absence of a net color field, and assuming isotropy and color
neutrality, the equilibrium distribution function for this system is,
 up to a normalization constant,
\begin{equation}
f^{(0)} (p) = \theta (\mu - p_0) \ ,
\label{eq:2.1}
\end{equation}
where $\theta$ is the step function, and $\mu$ is the chemical potential.
That is, all particle states are occupied with occupancy number one
up to the Fermi energy $p_0 = \mu$.

Now we study how the quark plasma reacts to  weak external color
electromagnetic fields. We consider
 that the chemical potential is large $\mu >>  \mu_c$, where $\mu_c$
is the critical chemical potential,
 while the color fields
are {\it soft} or of the scale order  $F_{\mu \nu} \sim g \mu^2$,
where $g$ is the coupling constant which is assumed to be small.
The distribution function can be expanded in powers of $g$ as:
\begin{equation}
f=f^{(0)}+gf^{(1)}+g^2f^{(2)}+...\ ,
\label{eq:2.2}
\end{equation}

 The Boltzmann equation (\ref{boltzmann}) for $f^{(1)}$ reduces to
\begin{equation}
p^{\mu} \left({\partial\over\partial x^{\mu}}-g\, f^{abc} A_{\mu}^b
Q_c {\partial\over\partial Q^a}\right)
f^{(1)}(x,p,Q) = p^{\mu} Q_a F_{\mu \nu}^a {\partial\over \partial
p_{\nu}} f^{(0)}(p_0)\ .
\label{eq:2.3}
\end{equation}
Notice that a complete linearization of the equation in $A_{\mu}^a$
would break the gauge invariance of the transport equation, which is
preserved in this approximation.
 But notice as well that this
approximation tells us that $f^{(1)}$ also carries a $g$-dependence.

 We can get the equation that the
 color current  density $J^{\mu}_a (x,p)$ obeys by multiplying
(\ref{eq:2.3}) by $p^{\mu}$ and $Q_a$ and  then integrating over the color
charges. For quarks in the fundamental representation
\begin{equation}
\int dQ \, Q_a Q_b = \frac{1}{2} \, \delta_{ab} \ .
\label{col}
\end{equation}
Taking also into account
\begin{equation}
\frac{d}{d p_0} \theta (\mu - p_0) = -\, \delta (\mu - p_0) \ ,
\label{der}
\end{equation}
we finally get, after summing over helicities,
\begin{equation}
[\,p \cdot D\,\, J^{\mu}(x,p)]^a = -g^ 2 \, p^{\mu} p^{\nu}
F_{\nu 0}^a (x) \, \delta (\mu -p_0) \ .
\label{eq:2.4}
\end{equation}
The total  induced current density can be obtained after
summing over the different contributions due to
different quark flavors.  In
order to simplify the notation we will not write the quark flavor
index in the color current, chemical potentials, masses, etc.,
and it should be understood that a sum over quark flavors
is to be taken in all final formulas.

{}From Eq. (\ref{eq:2.4}) we see that only quarks which are on the Fermi
surface contribute to the induced color current in the plasma. Quarks which
are inside the Fermi sea are blocked to react to the presence of
external fields due to the Pauli exclusion principle. Furthermore, the plasma
only responds to the presence of external color electric fields, and that
is why only those get a screening mass in this approach.

In order to solve (\ref{eq:2.4}), we first
divide the equation by $p_0$.
Then we integrate it over $p_0$ and $|{\bf p}|$, using the
momentum measure $dP$  (\ref{measurep}).
Due to the delta function in the momentum measure
and the delta function in the r.h.s of Eq. (\ref{eq:2.4}), $J^{\mu}_a (x,p)$ is
only non-vanishing when $p_0 = \sqrt{|{\bf p}|^2 + m^2} = \mu$.
Then we can write Eq. (\ref{eq:2.4}) as
\begin{equation}
[\, v\cdot D \ {\cal J}^{\mu}(x,v)]^a=- M^2
\, v^\mu\, v^\rho\, F^a _{\rho 0}(x)\ ,
\label{eq:2.6a}
\end{equation}
with
\begin{equation}
 M^2 = g^2  \frac{\mu p_F}{2 \pi^2} \ ,
\label{mass}
\end{equation}
where $p_F= \sqrt{\mu^2 -m^2}$ is the Fermi momentum
and $v^{\mu} = (1, \frac{p_F}{\mu} {\bf v})$,
and we have defined
\begin{equation}
{\cal J}_a ^\mu (x,v) = \int \frac{|{\bf p}|^2\, d|{\bf p}|\,dp_0}
{2 \pi^2}\,2\,\theta (p_0)\,
\delta(p^2-m^2)\,\,J_a ^\mu(x,p)\ ,
\label{eq:2.5}
\end{equation}
so that the total color current is obtained  by integrating over all
directions of the unit vector
 ${\bf v}$,
\begin{equation}
J^\mu _a (x) = \int \frac{d \Omega}{4 \pi} \, {\cal J}_a ^\mu (x,v) \ .
\label{intdir}
\end{equation}

 Equation (\ref{eq:2.6a}) has the same structure as  the transport
equation that the induced
color density quantity ${\cal J}_a^{\mu\, (T)} (x,V)$ in a
{\it softly} perturbed  hot quark-gluon plasma
obeys  \cite{KLLM}
\begin{equation}
[\, V \cdot D \ {\cal J}^{\mu\, (T)}(x, V)]^a=
- m_D ^2
\, V^\mu\, V^\rho\, F^a _{\rho 0}(x)\ ,
\label{eq:2.6b}
\end{equation}
where $m_D^2 = g^2 T^2 (N + N_f/2 )/3$ is the Debye mass squared
for a $SU(N)$ non-Abelian group, with $N_f$ flavors of quarks,
 and $V^{\mu}$ is a light-like four vector.

Equations (\ref{eq:2.6a}) and (\ref{eq:2.6b}) exhibit a similar structure,
although they correspond to two different physical situations.
In the case of a hot  quark-gluon plasma the contribution to the
induced color current comes from {\it both} quarks and gluons which
are thermalized at a temperature $T$. This is reflected in the
coefficient of the r.h.s. of (\ref{eq:2.6b}), the Debye mass squared.
For very high temperatures,
we expect that  quarks in the plasma move very fast, and
it is a good approximation to consider that their velocities
are ultrarelativistic, as long as $T \gg m$, where $m$ is the
quark mass, which is then neglected.
 Gluons travel at light velocities, however,
without any approximation. These are the reasons why the
vector $V^\mu$ in Eq. (\ref{eq:2.6b}) is a light-like four vector.
In the zero temperature
quark plasma that we are considering here there
are no {\it real} gluons, and that is why they do not contribute
to the induced color current. Furthermore, only quarks with
Fermi energy contribute to that current. For very large chemical potential,
it can be a good approximation to neglect quark masses if $\mu \gg m$.
If quark masses are neglected, then $p_F = \mu$, and
Eq. (\ref{eq:2.6a}) coincides with Eq. (\ref{eq:2.6b}), except for the
the factors $m_D^2$ and $M^2$.
We have decided, however, to keep corrections due to  quark masses,
which is justified if $\mu > m \gg g \mu$.  Therefore
$v^{\mu}$ is not a light-like vector in this more general situation
where we are considering massive quarks.

Before solving Eq. (\ref{eq:2.6a}) we would like to comment about
the connection between this transport problem and the Chern-Simons
eikonal.  Efraty and Nair \cite{EN1}, \cite{EN2}
 have shown that the gauge invariance
condition for the generating functional of hard thermal
loops $\Gamma_{HTL}$ of QCD can be formally related to the zero
curvature condition of a Chern-Simons
theory in a (2+1) dimensional space-time at zero temperature.
Furthermore, it has been shown that this  condition can be
obtained from the transport equation (\ref{eq:2.6b}) after assuming
$J^{\mu\, (T)}_a (x) =- \frac {\delta \Gamma_{HTL}[A]}{\delta A_{\mu}^a (x)}$.
The same identifications can be done for the cold quark  plasma,
after assuming that the induced color current can be obtained
from a generating functional. The same
 steps that were necessary to show that identification for the thermal
problem can be
repeated here with the only main difference that $v^{\mu}$ is not a
light-like vector, when quark masses are not neglected.

For completeness we present that identification below.
We first define a new current density
\begin{equation}
{\tilde {\cal J}}^{\mu a}(x,v) = {\cal J}^{\mu a}(x,v) + M^2 \,
 v^\mu A_0^a(x)\ .
\label{eq:2.7}
\end{equation}
Then ${\tilde {\cal J}}^{\mu a}$ obeys the equation
\begin{equation}
[ v\cdot D \  {\tilde {\cal J}}^{\mu}(x,v)]^a= M^2 \ v^\mu
{\partial\over\partial x^0} \,\Bigl(v\cdot A^a(x)\Bigr)\ .
\label{eq:2.8}
\end{equation}

Now we assume that ${\tilde {\cal J}}^{\mu}_a$
can be derived from a generating functional as
\begin{equation}
{\tilde {\cal J}}^{\mu}_a(x,v)={{\delta W(A,v)}\over
{\delta A_\mu ^a(x)}}\ .
\label{eq:2.9}
\end{equation}
Equation (\ref{eq:2.8}) then implies that $W(A,v)$ depends only on
 $A_+^a = \frac{1}{\sqrt{2 f}}  v\cdot A^a$, with
$  f = p_F / \mu$,
 {\it i.e.} $W(A,v) =W(A_+)$, and ${\tilde {\cal
J}}^{\mu}_a={{\delta W( A_+ )}\over{\delta A_+^a }}\,v^\mu$.
If we now define
 \begin{eqnarray}
 x_+ & = & \sqrt{\frac {f}{2}} \left( x_0 + \frac{1}{f} {\bf v} \cdot {\bf x}
\right)
 \  ,  \qquad  \partial_+  =   \frac{1}{\sqrt{2 f}}  \, v \cdot \partial  \ ,
\nonumber \\
 x_- & = & \sqrt{\frac {f}{2}} \left( x_0 - \frac{1}{f} {\bf v} \cdot {\bf x}
\right)  \ ,
\qquad
 \partial_- =  \frac{1}{\sqrt{2 f}}  \, {\bar v} \cdot \partial  \ ,
\label{eq:2.10}
\end{eqnarray}
with $\bar{v} \equiv (1,- \frac{p_F}{\mu} {\bf v})$,
and
\begin{equation}
F^a = {{\delta W( A_+ )}\over{\delta A_{+ a} }} - \frac12 M^2 A_+^a \ ,
\label{eq:2.11}
\end{equation}
then, after a Wick rotation to Euclidean space $x_+ \rightarrow z$,
$x_- \rightarrow {\bar z}$, $A_+ \rightarrow A_z$ and the
 identification
$F^a = - \sqrt{\frac{2}{f}}\frac{1}{M^2} a_{{\bar z}}^a$,
 Eq. (\ref{eq:2.8}) is translated into (in matrix notation, and using
antihermitian generators in the fundamental representation of $SU(N)$)
\begin{equation}
\partial_z a_{{\bar z}} - \partial_{{\bar z}} A_z + g [A_z,a_{{\bar z}}]=0 \ ,
\label{eq:2.12}
\end{equation}
which corresponds to the zero curvature condition of a Chern-Simons theory
in (2+1) Euclidean dimensional space-time, in the gauge $A_0 = 0$.

Solutions of Eq. (\ref{eq:2.12}) are provided by the Chern-Simons eikonal
and were studied long ago \cite{CS}. The generating functional of the
induced  color current in a cold quark plasma can then be expressed in terms of
the Chern-Simons eikonal, exactly as in the case of the hot quark-gluon plasma.

In spite of the very
interesting connection between these, in
principle, unrelated problems, we will not pursue the Chern-Simons approach
to the transport equations in the remaining part of this article.

\subsection{Solution to the transport equation}

Equation (\ref{eq:2.6a})
can be solved after inverting the $v \cdot D$
operator and  imposing proper boundary conditions.
Then the solution of (\ref{eq:2.6a}) is expressed in terms of
link operators \cite{TW}, \cite{BI}.
In this Subsection, we prefer to solve the transport equation by going to
momentum space exactly as has been done for Eq. (\ref{eq:2.6b}) in
ref.\cite{Zhang}.
Writing
\begin{equation}
{\cal J}_a ^{\mu}(k,v) = \int \frac{d^4 x}{(2 \pi)^4}\, e ^{i k \cdot x}\,
 {\cal J}_a ^{\mu}(x,v) \ ,
\label{eq:3.1}
\end{equation}
Eq. (\ref{eq:2.6a}) becomes in momentum space
\begin{eqnarray}
& &v \cdot k \, {\cal J}^{\mu} _a(k,v) +
i g f_{abc} \int \frac{d^4 q}{(2 \pi)^4} \,
v \cdot A^b (k-q)\, {\cal J}^{\mu c} (q,v) \nonumber \\
& & =  - M^2 v^{\mu} \left[ v \cdot k \, A_0 ^a (k) - k_0\, v \cdot A^a (k) +
i  g f_{abc} \int \frac{d^4 q}{(2 \pi)^4} \,
v \cdot A^b (k-q) A_0 ^c (q) \right] \,
\label{eq:3.2}
\end{eqnarray}
Now, after assuming that ${\cal J}_a^{\mu} (k,v)$ can be expressed as an
infinite power
series in the gauge field $A^a _{\mu} (k)$, Eq. (\ref{eq:3.2}) can be
solved iteratively for each order in the power series.
We will impose retarded boundary conditions by the prescription
$p_0 \rightarrow p_0 + i \epsilon$, with $\epsilon \rightarrow 0^+$,
 that should be understood in all
following formulas.

The first order solution is
\begin{equation}
{\cal J}^{\mu\, (1)}_a (k,v) =
  M^2 \, v^{\mu}
\left (k_0 \, \frac{ v \cdot A_a (k)} {v \cdot k}- A^0 _a (k) \right)
\ .
\label{cur1}
\end{equation}
Inserting (\ref{cur1}) in (\ref{eq:3.2}) allows solving for
the second order term in the series, which reads
\begin{equation}
{\cal J}^{\mu \,(2)}_a (k,v) =
-i g  M^2 f_{abc}  \int \frac{d^4 q}{(2 \pi)^4} \, v^{\mu} q_0
 \frac{ v \cdot A^b (k-q) \, v \cdot A^c (q)}{(v \cdot k) (v \cdot q)}
\ .
\label{cur2}
\end{equation}
The $n$-th order term ($n>2$) can be expressed as a function of the
$(n-1)$-th one as
\begin{equation}
{\cal J}^{\mu \,(n)}_a (k,v)=
 -i g f_{abc} \int \frac{d^4 q}{(2 \pi)^4}  \,
 \frac{ v \cdot A^b (k-q)} {v \cdot k} \, {\cal J}^{\mu \,(n-1)}_c (q,v)
\ .
\label{curn}
\end{equation}

The complete expression of the induced color current in the cold quark
plasma is thus given by
\begin{eqnarray}
J^{\mu}_a (x) &=& \int \frac{d^4 k}{(2 \pi)^4}\, e ^{-i k \cdot x}\,
\sum_{n=1}^{\infty} J_a ^{\mu \,(n)}(k) \nonumber \\
&=& \int \frac{d \Omega}{4 \pi} \int \frac{d^4 k}{(2 \pi)^4}\, e ^{-i k \cdot
x}\,
\sum_{n=1}^{\infty} {\cal J}_a ^{\mu \, (n)}(x,v) \ .
\label{currto}
\end{eqnarray}
The color current is then expressed as an average over
the three dimensional unit vector ${\bf v}$,
which represents the directions of the velocity
vectors of quarks with Fermi energy.

It should be possible to construct a generating functional $\Gamma_{HDL}$ that
generates this current by solving
\begin{equation}
J^{\mu}_a (x) = - \frac{\delta \Gamma_{HDL} [ A ]} { \delta A_{\mu}^a (x)} \ .
\label{gener}
\end{equation}
It is first required to check that
 $J^{\mu}_a (x)$ obeys integrability conditions, as has been done
 in ref. \cite{TW} for the massless and thermal case.

\subsection{Polarization Tensor}

The  polarization tensor $\Pi^{\mu\nu}_{a b}$ can be computed from
(\ref{cur1}) by using the relation
\begin{equation}
J^{\mu \, (1)}_a (k) =  \Pi ^{\mu\nu} _{a b} ( k)\,
A_{\nu} ^{b} (k) \ .
\end{equation}
It reads:
\begin{equation}
\Pi^{\mu\nu}_{ab} ( k)=  M ^2 \left (-g^{\mu 0} g^{\nu 0} +
k_0 \, \int \frac{d \Omega}{4 \pi} \,
\frac {v^{\mu} v^{\nu}} {v \cdot k}  \right)  \,
\delta_{ab} \ ,
\label{polar}
\end{equation}
where we recall that $v^{\mu} = (1, \frac{p_F}{\mu} {\bf v})$.
To avoid the poles in the above  integrand, we impose retarded
boundary conditions, {\it i.e.}, we replace $k_0$ by  $k_0 + i
\epsilon$.

The  polarization tensor (\ref{polar})  obeys the Ward
identity
\begin{equation}
k_{\mu} \Pi^{\mu\nu}_{ab} = 0 \ ,
\end{equation}
and it is also gauge-independent due to the gauge invariance of
the transport formalism.

The real  part of the polarization tensor is
\begin{mathletters}
\begin{eqnarray}
{\rm Re} \, \Pi^{00}_{ab} (k_0, {\bf k}) & = &\delta_{ab} \, \Pi_{l}
(k_0, {\bf k}) \ ,\\[2mm]
{\rm Re} \, \Pi^{0i}_{ab} (k_0, {\bf k}) & = & \delta_{ab} \,
k_0 \, \frac{k^i}{|{\bf k}|^2}\, \Pi_{l} (k_0, {\bf k}) \ , \\
{\rm Re} \, \Pi^{ij}_{ab} (k_0, {\bf k}) & = & \delta_{ab} \left[
 \left ( \delta^{ij}- \frac{k^i k^j}{|{\bf k}|^2} \right) \Pi_{t} (k_0,
{\bf k})+ \frac{k^i k^j} {|{\bf k}|^2} \, \frac{k_0^2}{|{\bf k}|^2} \,
\Pi_{l} (k_0, {\bf k}) \right] \ ,
\end{eqnarray}
\label{resultreal}
\end{mathletters}
$\!\!$where
\begin{mathletters}
\begin{eqnarray}
\Pi_{l} (k_0, {\bf k}) & = & M^2 \left( \frac{\mu}{p_F} \frac{k_0}{2|{\bf
k}|} \,{\rm ln\,}\left|{\frac{  \mu \, k_0+ p_F |{\bf k}|}
{ \mu \, k_0 -p_F |{\bf k}|}}\right|
-1  \right) \ , \\
 \Pi_{t} (k_0, {\bf k}) & = &-  M^2 \, \frac{k_0^2}{|{\bf
k}|^2} \left[ 1 + \frac12 \left( \frac{p_F}{\mu} \frac{|{\bf k}|}{k_0}-
\frac{\mu}{p_F}
\frac{k_0}{|{\bf k}|} \right) \,  {\rm ln\,} \left|{\frac{ \mu \, k_0+
 p_F |{\bf k}|}{\mu \,  k_0-p_F |{\bf k}|}}\right| \, \right] \ .
\end{eqnarray}
\label{pipi}
\end{mathletters}
The imaginary part of the polarization tensor
describes damping in the quark plasma,  explicitly:
\begin{equation}
{\rm Im} \, \Pi^{\mu\nu}_{ab} (k_0 ,{\bf k})  =  - \delta_{ab}\
M^2 \,\pi\, k_0 \int \frac{d \Omega}{4 \pi}\, v^{\mu}\, v^{\nu}\,
\delta(k_0 -  \frac{p_F}{\mu} \bf{k} \cdot \bf{v}) \ .
\label{landaudamp}
\end{equation}

Our result of the polarization tensor agrees with the one computed in
ref. \cite{Heinz} using classical transport theory for a quark-gluon
plasma at finite temperature and
baryon density in  the zero temperature limit.

Notice that if quark masses are  neglected, then $p_F = \mu$, and
then (\ref{polar}) reduces to the same polarization tensor as the one
found in a hot quark-gluon plasma \cite{EH},
 \cite{Heinz},\cite{KLLM}, with a different
screening mass.

Higher order $n$-point functions can be found as
\begin{equation}
\delta^{(4)} ( k -\sum_{i=1} ^{i=n-1} q_i) \,
 \Gamma^{\mu \nu_1 \cdots \nu_{n-1}}
_{a b_1  \cdots b_{n-1}} (k, q_1, \cdots, q_{n-1} ) =
\left.
\frac { \delta J^{\mu} _a (k)}{ \delta A_{\nu_1} ^{b_1} (q_1)
 \cdots \delta A_{\nu_{n-1}} ^{b_{n-1}} (q_{n-1})}
\right \vert_{A_{\rho} ^c = 0} \ ,
\label{npoint}
\end{equation}
where the delta function on the r.h.s. of (\ref{npoint}) accounts for
conservation of momentum, and permutations of indices and momentum
have to be taken.

Due to the fact that the induced color current can be expressed as an
infinite power series in the gauge field $A_{\mu}^a$, it is obvious that
there are $n$-point gluonic amplitudes, with arbitrary $n$, from 2 to
$\infty$. All these gluonic amplitudes describe the polarizability
properties of the non-Abelian plasma.

\section{THE POLARIZATION TENSOR OF QED AT FINITE DENSITY}
\label{sec5}

In this Section we present the computation of the polarization tensor
of QED due to the presence of a background density of electrons,
 and from it we
extract the HDL corresponding to that amplitude in the massless case.  This
computation has been previously performed by several authors in
Euclidean space-time, see ref. \cite{FMc}, \cite{Ka}, \cite{To},
\cite{Shur},
 and also \cite{AGD} and \cite{Fetter} for the non-relativistic
case.
 Here we review how this computation is performed in
Minkowski space-time when retarded boundary conditions are imposed.
 We find it more appropriate to carry out this computation in Minkowski
space-time to obtain the correct analytic properties
of the retarded polarization tensor.

In the presence of a finite density of electrons
the usual definition of creation and annihilation operators
for particles and ``holes", or antiparticles, of the vacuum
has to be modified \cite{Fetter}.
The ground state $|\Phi_0>$ of the
system is constituted by the Fermi sea, that is, by electrons occupying all
particle states according to the Pauli exclusion principle up to the
Fermi energy $p_0 = \mu$. Then one defines a
creation operator $b^{\dag}_{\alpha} (p)$ that acting on the ground state
creates one particle with energy $p_0 > \mu$, and the creation operator
$d^{\dag}_{\alpha} (-p)$, for $p_0 < \mu$, which
 creates a ``hole", which may be interpreted as the
absence of one electron in the Fermi sea. The corresponding
particle and hole destruction operators annihilate the ground state
\begin{equation}
 b_{\alpha} (p) \, |\Phi_0> = d_{\alpha} (p) \, |\Phi_0> = 0 \ ,
\end{equation}
since the ground state doesn't contain electrons with energies above
the Fermi energy or ``holes" inside the Fermi sea.

The usual averages of products of creation and annihilation
operators that are required to compute Feynman amplitudes
are then modified from their vacuum values,  explicitly,
\begin{equation}
< \Phi_0|\, b_{\alpha} (p) b_{\alpha'}^{\dag} (p)\, |\Phi_0>  =
  \theta (p_0-\mu) \, \delta_{\alpha, \alpha'} \ .
\label{averages}
\end{equation}
Then one can compute the Feynman propagator to get
\cite{Fetter}, \cite{Nair}
\begin{eqnarray}
\label{propagator}
i S_F (x,y)  & = & <\Phi_0 | {\cal T} \psi(x) {\bar \psi} (y) |\Phi_0>
\\
        & = &  \theta(x^0 - y^0) \,\int \frac{d^3 p}{(2 \pi)^3}\,
 \frac{1}{2 p_0}
\left[ (\gamma \cdot p +m) \,  \theta (p_0 - \mu)
\,  e^{-i p(x-y)}
\right ]
\nonumber \\
 & - &
\theta(y^0 - x^0) \, \int \frac{d^3 p}{(2 \pi)^3}\,
 \frac{1}{2 p_0}
\left[ (\gamma \cdot p+ m) \, \theta (\mu-p_0) \, e^{-i p(x-y)} +
(\gamma \cdot p - m) \, e^{i p(x-y)} \right] \nonumber
\end{eqnarray}
where $p_0 = \sqrt{ |{\bf p}|^2 + m^2}$. Notice that in the limit
$\mu \rightarrow 0$, (\ref{propagator}) agrees with  the Feynman propagator
in the vacuum.

In order to compute the
polarization tensor of QED in the presence of a background
 density of electrons,
the propagator (\ref{propagator}) is needed.
Retarded boundary conditions are taken into account by introducing
convergence factors $e^{\epsilon x_0}$, $e^{-\epsilon y_0}$, with
$\epsilon \rightarrow 0^+$,
 as needed when going from configuration to
momentum space (see ref. \cite{EN2} and \cite{JN}
for explicit details). Taking that into account,
we find for the chemical potential dependent
part of  the retarded polarization tensor in the massless case
\begin{eqnarray}
&\Pi& ^{\mu \nu} (k)  =   g^2 \int \frac{d^3 q}{(2 \pi)^3}
\frac{1}{2 p_0} \frac{1}{2 q_0} \left[
 (\theta (\mu-q_0) - \theta (\mu-p_0)) \frac{T^{\mu \nu} (p,q)}
{p_0 - q_0 - k_0 - i \epsilon} \right. \\
&-& \left.
 \theta (\mu - p_0) \frac{T^{\mu \nu} (p, q')}{p_0 + q_0 - k_0 -i \epsilon}
- \theta (\mu - q_0) \frac{T^{\mu \nu} (p',q)}{p_0 + q_0 + k_0 + i \epsilon}
\right]
\end{eqnarray}
where  $p= (p_0 , {\bf p})$, $p' = (p_0, -{\bf p})$,
$q = (q_0, {\bf q})$, $q' = (q_0, -{\bf q})$, with $p_0 = |{\bf p}|$,
 $q_0 = |{\bf q}|$,
 and
${\bf p} = {\bf q} + {\bf k}$. We have further defined
\begin{equation}
T^{\mu \nu} (p,q) \equiv {\rm Tr} \left [\gamma^{\mu} \gamma \cdot p
 \gamma^{\nu} \gamma \cdot q  \right] \ .
\end{equation}

In  the high chemical potential limit, one considers $k_0, |{\bf k}| \ll \mu$.
 Then one can make the approximations
\cite{FT}, \cite{Nair}
\begin{eqnarray}
T^{\mu \nu} (p,q) &\sim & 8 q_0 ^2 V^{\mu} V^{\nu} \\
T^{\mu \nu} (p, q') \sim T^{\mu \nu} (p', q) &\sim& 4 q_0^2
\left( V^{\mu} {\bar V}^{\nu} + {\bar V}^{\mu} V^{\nu} - 2 g^{\mu \nu} \right)
\\
p_0 - q_0 - k_0   &\sim & - k \cdot V \\
p_0 + q_0 \pm k_0 &\sim & 2 q_0 \\
\theta (\mu- q_0) - \theta (\mu- p_0) &\sim&
{\bf V} \cdot {\bf k} \, \delta (\mu - q_0) \ ,
\label{approx}
\end{eqnarray}
where $V=( 1 ,\frac{\bf q}{q_0})$, ${\bar V} = (1, - \frac{\bf q}{q_0})$
to finally get
\begin{equation}
\Pi^{\mu\nu} ( k)= - g^2 \int \frac{d ^3 q}{(2 \pi)^3}
\left ({\bar V}^{\mu} V^{\nu} + V^{\mu} V^{\nu}
-2 \,k_0 \, \frac {V^{\mu} V^{\nu}} {V \cdot k +
i \epsilon} \right)
\delta(\mu-  q_0)  \ ,
\end{equation}
which after performing the integral over the modulus of $|{\bf q}|$
reduces to the same polarization tensor (\ref{polar}) for the Abelian
case and massless case.

\section{CONCLUSIONS}
\label{sec6}

In this paper we have used classical transport theory to study the
response of a completely degenerate non-Abelian plasma at zero
temperature to weak color electromagnetic fields.
We have used an approximation scheme for the classical
transport equations that respects their non-Abelian gauge symmetry,
and from it we have obtained the induced color current in the
non-Abelian plasma.

The  study of the transport phenomena occurring at high density
in a quark plasma that we have carried out here is
quite similar to the one that was done in
\cite{KLLM} for a QCD plasma at high temperature, and we direct the
reader to consult that reference.
Throughout this paper we have stressed the parallelism that these
two different problems exhibit, as well as their differences.

 Both at high temperature and
high density, the color constituents of the non-Abelian plasma
are not confined, and that is why one can model color degrees
of freedom classically.

The transport equations obeyed by the induced color current for a non-Abelian
plasma at high temperature and at high density  present a similar structure.
 If quark masses are neglected, then the two transport
equations are essentially the same, except for  the following.
 At high
temperature $T$, both quarks and gluons which are thermalized at the
temperature $T$ contribute to the induced color current in the plasma.
At high chemical potential, only quarks on the Fermi surface
contribute to that current, as there are no {\it real} gluons at zero
temperature. Then, in that situation,
the only difference between the two transport equations and their solutions
comes from the screening masses squared which are
$m_D^2 = g^2 T^2 (N + N_f/2 )/3$
and $M^2 = g^2 N_f \mu^2/ 2\pi^2$, respectively,
for a $SU(N)$ theory with
$N_f$ flavors of quarks (and assuming that for all quark flavors
$\mu_f = \mu$).

We have decided to keep corrections due to quark masses in the
 present study. The solution to the transport equation can
be expressed in terms of an average of a three-dimensional
unit vector ${\bf v}$, which represents the directions of the
velocity vectors of quarks which have  Fermi energy.

We have also shown that the solution of the transport equation
can be expressed in terms of the Chern-Simons eikonal,
even for massive quarks.
Even though one cannot define natural
light-cone coordinates when quarks are not consider ultrarelativistic,
we have seen that it is still possible to relate the transport
equation to the zero-curvature condition of a Chern-Simons theory
in (2+1) Euclidean space-time. The transport equation only depends
on the projections of the vector gauge fields over
 the four vectors $v^{\mu}$ and ${\bar v}^{\mu}$
defined in Subsection III~A, and this is enough to define a two dimensional
plane where the zero-curvature condition of a Chern-Simons theory
can be recognized, after proper identifications.

We have presented a systematic way of computing $n$-point gluonic
amplitudes using the solution of the classical
transport equation, and we claim
that these results should  agree with the high density limit
of a quantum field theory approach. Complete agreement is found for
the polarization tensor ($n=2$) in the massless case,
 and complete agreement should be expected for
higher $n$-point functions, although we have not check this statement,
not are we  aware of the computation of
those amplitudes in the literature.

Therefore, we claim that a Braaten-Pisarski resummation procedure
should also be used in QCD with high chemical potential, as  was
generally expected, although a more detailed quantum field theoretical
study should be carried out to see exactly how this resummation is implemented.
We think that it is interesting to have this alternative computation
of HDLs, specially due to the
simplicity and transparent physical interpretation of the classical
transport formalism.

Finally, we should mention that although the response theory
of the  high temperature limit
(at zero chemical potential) and the high chemical potential limit
(at zero temperature) of QCD  seem almost identical,
and one could naively conclude that
in both situations static color electric fields are exponentially
damped (at least at leading order), this is not so.
Kapusta and Toimela \cite{KapTo} have pointed out that the
static potential between two static charges in a plasma at
zero temperature becomes oscillatory and vanishes as a power
at large distances due to the existence of a sharp Fermi surface.
This effect was known as Friedel oscillation in non-relativistic
quantum theory \cite{Fetter}.

A further study of the high density limit of QCD should  be
carried out, and we postpone that study to a further publication.

\vspace{20mm}
{\bf Acknowledgements:}

\vspace{5mm}
I am especially grateful to E.~Braaten, for suggesting me the
interest of pursuing this study, and to R.~Jackiw, for very
enlightening discussions.
I also want to thank the comments on an early
version of this manuscript of D.J.~Casta\~no, F. ~Guerin and C.~Lucchesi,
as well as instructive e-mail discussions with Y.~ Lozano.

This work is supported in part by funds provided by the
Ministerio de Educaci\'on y Ciencia, Spain, through a FPI
fellowship and by
U.S.~Department of Energy (D.O.E.) under cooperative agreement
\#~DE-FC02-94ER40818.

\end{document}